\documentclass[preprint, authoryear, 5p, number]{elsarticle}
\usepackage{hyperref}
\usepackage{units}
\usepackage{amssymb}
\usepackage[english]{babel}
\usepackage{color}
\usepackage{array}
\usepackage{booktabs}
\usepackage{graphicx}
\usepackage{array}
\usepackage{subfig}
\usepackage{fnpos}
\usepackage[T1]{fontenc}
\hypersetup{
    colorlinks=true,
    linkcolor=blue,
    filecolor=magenta,      
    urlcolor=blue,
}

\newcolumntype{M}[1]{>{\centering\arraybackslash}m{#1}}
\newcolumntype{N}{@{}m{0pt}@{}}

\journal{Publications of the Astronomical Society of the Pacific}

\begin{document}

\newcommand{\HI}{H\textsc{i}}
\newcommand{\Msolar}{M$_{\odot}$}
\newcommand{\kms}{km\,s$^{-1}$}

\begin{frontmatter}

\title{Visualising three-dimensional volumetric data with an arbitrary coordinate system}

\author[a]{R. Taylor\corref{cor1}}
\ead{rhysyt@gmail.com}
\cortext[cor1]{Corresponding author}
\address[a]{Astronomical Institute of the Czech Academy of Sciences, Prague}

\begin{abstract}
Astronomical data does not always use Cartesian coordinates. Both all-sky observational data and simulations of rotationally symmetric systems, such as accretion and protoplanetary discs, may use spherical polar or other coordinate systems. Standard displays rely on Cartesian coordinates, but converting non-Cartesian data into Cartesian format causes distortion of the data and loss of detail. I here demonstrate a method using standard techniques from computer graphics that avoids these problems with 3D data in arbitrary coordinate systems. The method adds minimum computational cost to the display process and is suitable for both realtime, interactive content and producing fixed rendered images and videos. Proof-of-concept code is provided which works for data in spherical polar coordinates.
\end{abstract}


\end{frontmatter}

\section{Introduction}
\noindent As computing resources and observational equipment has improved, three-dimensional data sets have played an increasingly important part in understanding astronomy. Using multibeam instruments, neutral atomic hydrogen (\HI{}) surveys have created megapixel maps with thousands of velocity channels (e.g. ALFALFA, \citealt{aa}), with SKA pathfinder telescopes expected to create much larger data sets (e.g. WALLABY, \citealt{kor}) and increase the number of detections by several orders of magnitude (\citealt{P15}). Recent developments at optical wavelengths have seen a great increase in the number of Integral Field Units, offering millions of spectra in a field of view $\sim$1 arcminute$^{2}$ (e.g. \citealt{muse}). Numerical simulations routinely use 3D analysis for both particle and grid-based codes.

There are two approaches to dealing with this increasing quantity of multi-dimensional data. The first is to develop more sophisticated algorithms, increasingly relying on automatic methods to at least identify features of interest and possibly measure them as well (see \citealt{floer}). The second approach is to develop better tools for visualisation (see \citealt{P15}) that allow users to more easily manipulate and analyse data. To some extent the first approach is both necessary and desirable : visual analysis is subjective and time consuming for large data sets. But algorithms cannot produce meaningful output if the data is too different from expectations, so they can miss or incorrectly process the most interesting discoveries.

In \cite{me15} (hereafter T15, see also \citealt{me14}) I described \textsc{frelled}, a tool designed to address some of the problems of visual analysis. Specifically, for \HI{} data cubes one major problem has been that while an observer can almost instantly identify a source in the data set, recording the position and parameters of the source can be very much slower. The user has had to rely on memory to determine if a source has been catalogued already, with no easy way to mask the data. \textsc{Frelled} allows the user to interactively define 3D regions which can simultaneously mask data and record source parameters. This was shown to give an increase in visual cataloguing speed by as much as a factor of 50 (although this does depend on user experience, data size and source density, it was also shown that using \textsc{frelled} is almost always faster than other available tools). Data sets that previously took weeks to analyse can now be catalogued in hours or days.

\textsc{Frelled} can process any 3D data set in the standard FITS format. It allows the user to view volumetric data in realtime, freely rotating the view without loss of detail as the view is recalculated. This is particularly useful for numerical simulations where the data are in true 3D coordinates (rather than position-velocity space as for \HI{} and other observational data) since projection effects can be easily checked.

\textsc{Frelled} is a series of Python scripts for the open-source graphics suite `Blender', so it has to operate within the restrictions of Blender's available features. While it is relatively straightforward to display data in Cartesian coordinates, other coordinate systems are more difficult. The remainder of this paper is organised as follows. In section \ref{sec:meth} I describe the pitfalls of trying to convert between non-Cartesian and Cartesian systems and describe a method to circumvent these problems. In section \ref{sec:res} I demonstrate the results of this using two example data sets. In section \ref{sec:disc} I discuss the advantages and disadvantages of this and outline how this could be developed into a working visualisation tool. As in T15, all visualisation was performed on an HP Elite 7500 series desktop with an Intel i7-3770 quad core 3.4 GHz CPU, 16 GB RAM and a 4GB NVIDIA GeForce GT 640 GPU. The version of Blender used is 2.49b which uses Python 2.6.

\section{Methods}
\label{sec:meth}
\subsection{Displaying volumetric data in Blender}
\label{sec:blend}
Blender is primarily a mesh-modelling tool in which users create virtual objects for use in static images, animations, and interactive content. An object in Blender can hold several different sorts of information : a virtual camera from which to render the viewpoint, a lamp to create light and shadows, mathematical curves and surfaces, or other constructs used to manipulate the virtual objects. The main type of object is a \textit{mesh}, which consists of a set of vertices which define faces. Meshes can be assigned various sorts of materials : halos can display individual vertices as points, wireframe materials render only the edges of the faces which link vertices. Standard materials render the faces with different colours, transparency, reflectivity and other effects.

Blender is not optimised to deal with very large data sets, so importing a standard FITS file (which might contain many millions of data points) as individual Blender objects is not feasible. The technique used in \textsc{frelled} is to slice the data, assuming it to have a simple Cartesian grid coordinate system. Each slice is rendered as a PNG image file, where the user controls the image parameters - e.g. the data range and colour scheme. The image slices are displayed in Blender on plane meshes which consist of four vertices which define a single face. The images are displayed as a texture on the Blender material assigned to each mesh (see T15 figure 1). This minimizes the number of vertices (and so memory usage) that Blender requires. The texture controls not only the colour of the material but also its opacity\,: thus denser regions are more opaque and lower-density regions appear partially or entirely transparent. In this way it is possible to display data cubes equivalent to $\sim$512$^{3}$ voxels, depending on the capabilities of the computer (especially the memory capacity of the GPU), at $>$ 15 f.p.s (frames per second).

This approach allows for considerable control over the display. For example, multiple FITS files can be displayed simultaneously with different colour transfer functions\footnote{The transfer function defines how the data values are converted into the RGB values on the screen.} (multi-volume rendering). Different files can be used to control the colour and opacity. Particles and (velocity) vectors can be easily overlaid using other mesh objects, and the user can interactively define regions on which to do basic analysis tasks - for example, querying external databases, producing contour maps, and for \HI{} data one can use the regions to run the spectral analysis task \textsc{miriad} (\citealt{miriad}) with an interactive GUI. The features are described in full in T15.

\subsection{The problems of displaying non-Cartesian data}
\label{sec:carts}
\noindent In a Cartesian system one knows the $x,y$ coordinates of a data value $v$. Assuming the pixels are equally sized it is straightforward to construct an image - a colour transfer function is used to transform $v$ into some arbitrary RGB values at each point on a plane, it being trivial to define the boundaries of each pixel since they are rectangular. Thus it is simple to define the displayed colour at any point since the colour values are defined at every $x,y$ point. Problems arise when one starts with data which are in non-Cartesian systems, for example in polar coordinates.

\begin{figure*}[t]
\begin{center}
\includegraphics[width=185mm]{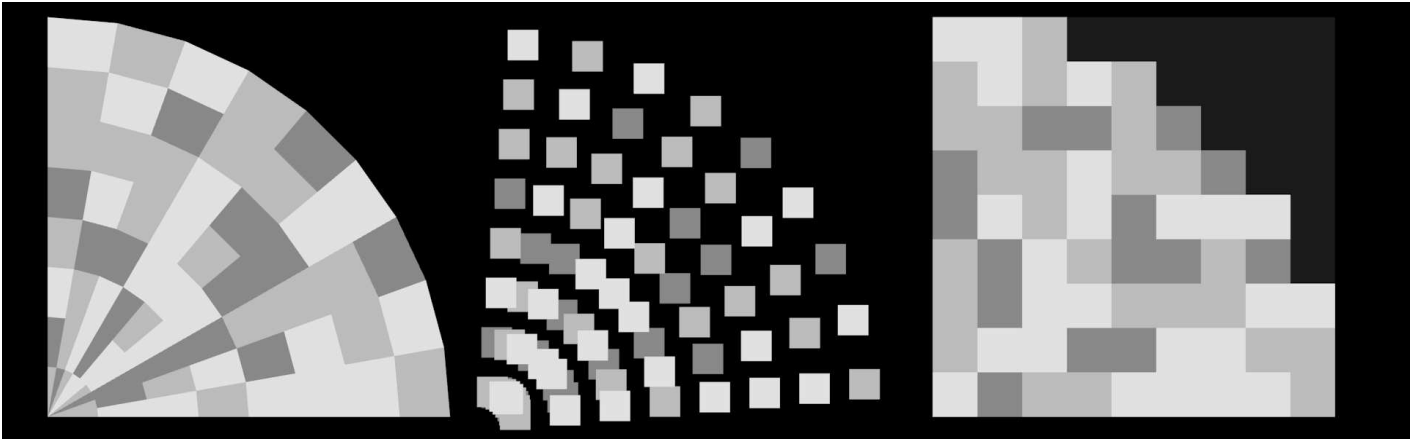}
\caption[Mapping]{Problems of converting polar coordinate data into Cartesian images. The left panel shows the data in its native polar coordinate system. The middle panel shows what happens if the centre of each data cell is used to set the $x,y$ position of a pixel of fixed size (or in Blender one may generate an object). Towards the centre the pixels strongly overlap and information is lost, while at larger distance the image is underfilled - the angular separation of the original cells is constant, so their separation in Cartesian coordinates varies with radius. The right panel uses a Cartesian grid, setting each pixel value based on the polar cell at the coordinates of the pixel centre. Although the data is now fully sampled, there is again loss of detail near the centre and distortions at larger radii.}
\label{fig:mapping}
\end{center}
\end{figure*}

With polar coordinate data, one knows the $\phi,r$ coordinates of the value $v$. The problem is that that each $\phi,r$ value may not specify every $x,y$ position on the Cartesian grid used to generate the image, as illustrated in figure \ref{fig:mapping}. It is usually trivial to find the $x,y$ coordinates of every $\phi,r$ position, but it is less obvious how to check which cell any $x,y$ position corresponds to in the original polar coordinate data. Without accounting for this, the $x,y$ grid can be underfilled at large $r$ while at low $r$ detail from the original data will be lost, since many $\phi,r$ points can correspond to the same $x,y$ position. Even in the best case where the Cartesian data is fully sampled, the original data is distorted and detail is lost.

The solution I present here is to display the data in its native coordinate system. The most obvious method would be to generate a Blender mesh where the vertices define faces using the polar coordinate system, since this is no more difficult than using the Cartesian system\footnote{Basic code to do this is described and available for download at the following URL : \href{http://www.rhysy.net/fits-method-2-1.html}{http://www.rhysy.net/fits-method-2-1.html}}. This does not necessarily mean generating a single Blender object per pixel - instead a single object could be generated comprised of many faces and materials. Blender is much more optimised to handle large numbers of faces than it is large numbers of objects. However, with millions of faces and materials this would still give a prohibitively slow performance. The solution, as with \textsc{frelled}, is to use image textures, but now there is the added complication of correcting for the distortion.

The distortion correction can be done directly in Blender (or indeed any standard CGI modelling software) by the technique known as \textit{UV mapping}. This maps an image to a mesh by specifying how the coordinates of the image relate to the coordinates of the mesh vertices. If the mesh vertices are moved, the image texture is `deformed' to follow the new coordinates. The image is stretched via linear interpolation between the vertices.

The technique here is to initially pretend the data is in Cartesian coordinates, which makes it trivial to convert the data into standard image files that Blender can read - specifically PNGs which are generated with standard \textit{matplotlib} routines in Python (see T15). For example if the data is really in spherical polar coordinates, the PNG image generated can use the $\phi$ coordinate for its x-axis and the $r$ coordinate for its y-axis. This means the data is at all times fully sampled : no detail is lost at any position. For 3D data the image will be a slice through the data at a constant value of $\theta$. Initially, the PNG is mapped onto a planar grid mesh in Blender, which makes generating the UV map extremely simple : the position of each point on the mesh is directly equivalent to its UV coordinates. For 3D data there will be a series of plane meshes, each at a constant $z$ position in Blender corresponding to the $\theta$ position of the original data slice.

With the UV map created, the mesh vertices can then be moved via Python script into their correct positions. The script extracts their polar coordinates from their initial positions and applies the standard equations ($x\,=\,r\,sin(\theta)\,cos(\phi)$; $y\,=\,r\,sin(\theta)\,sin(\phi)$; $z\,=\,r\,cos(\theta)$) to set their correct Cartesian coordinates. An animation of this process is available at the following URL :

\noindent \href{https://youtu.be/8CQi9TFvJ90}{https://youtu.be/8CQi9TFvJ90}.

In this simple way, the problems of losing detail when displaying the data are completely eliminated. It does not depend on the data being in polar coordinates - any coordinate system can be used. All one needs to know is the equations that relate the original grid and Cartesian systems, or even a lookup table if the relationship is nonlinear. The original grid can either be generated via Blender's tools or via a Python script. Whereas for displaying volumetric Cartesian data a series of plane meshes are used to display slices of the data, with this technique polar data is displayed on a series of cones or spheres, each one corresponding to a constant $\theta$ or $r$ value.

One disadvantage of this method is that to correct for the coordinate transform with perfect accuracy, an equal number of vertices as the number of data cells (voxels) is required. A one megapixel image in Cartesian coordinates can be displayed in Blender with four vertices at $>$ 60 f.p.s. on a standard desktop PC. If this was polar coordinate data that single image would require one million vertices, which drops the frame rate to $<$ 1 f.p.s. For 3D data tens or hundreds of such images may be required, which would obviously make this technique unviable.

\begin{figure*}[t]
\begin{center}
\includegraphics[width=185mm]{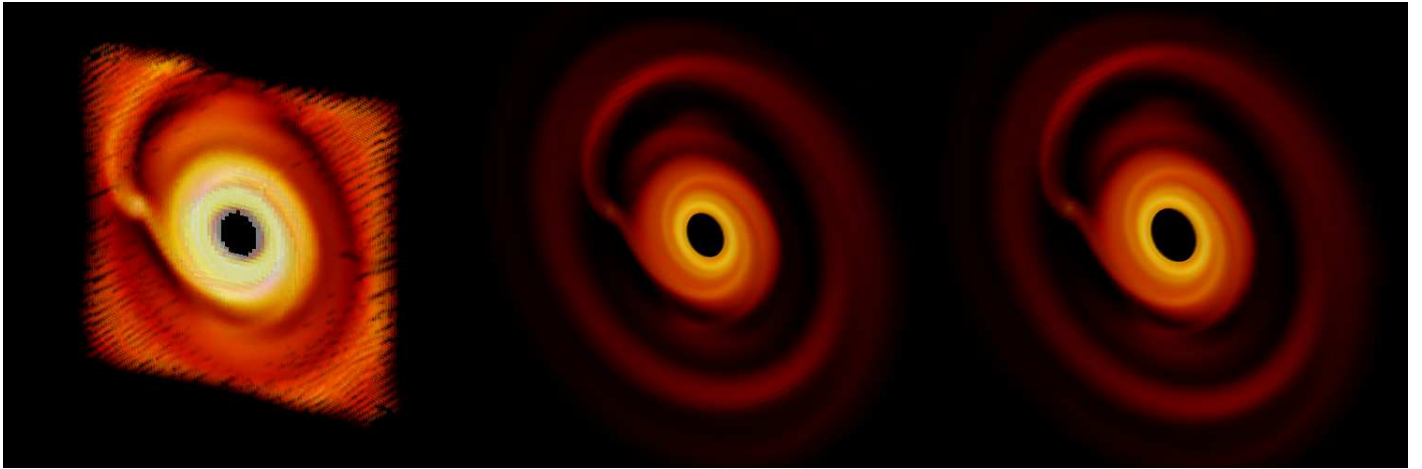}
\caption[Conversion]{Practical examples of the different techniques used to render data in spherical polar coordinates, from a protoplanetary disc simulation (Hubert Klahr private communication but see \citealt{klahr} for a description of the code). In the left panel the data was gridded into Cartesian coordinates while in the others it was rendered directly using spherical polar coordinates as described in the text. The middle panel used a mesh resolution equal to the data resolution, giving a vertex count of 1.9 million. The right panel used a lower mesh resolution giving a vertex count of 52,000. This particular data set has a variable radial pixel scale so a lookup table was used to stretch the radial pixel coordinates to the correct position. If the radial pixel scale was constant, an even more dramatic reduction in the vertex count could be achieved, to as low as 6,500. Rotating movies can be seen at the following URL : \href{https://youtu.be/Pa0Fwu0YZ9c}{https://youtu.be/Pa0Fwu0YZ9c}.}
\label{fig:conv}
\end{center}
\end{figure*}

Fortunately there are several ways to overcome this problem. The simplest method is not to use one vertex per pixel and accept a distortion between pixels. The data itself can make this easier - if the size of each pixel is constant in the $r$ direction, there is no need for one vertex per radial pixel - Blender's linear interpolation of the texture will give exactly the same result as if it was UV mapped to a series of vertices. This would reduce the million vertices required for a one megapixel image to a mere one thousand. Allowing a slight distortion, the number of pixels in the $\theta$ direction can be reduced by a factor of a few. Thus instead of a million vertices per image, only hundreds are required. A comparison of the renders using different mesh resolutions is shown in figure \ref{fig:conv}.

More drastic options are also available. Blender's game engine offers considerably superior performance to its realtime display, giving $\geq$ 50 f.p.s. even with one million vertices. The disadvantage is that this does not have the same navigation interface as in the standard view, though one could be designed (for a discussion on why this is extremely valuable for data analysis, see T15). A more extreme option would be to employ a different rendering engine altogether, with other software enabling high frame rates with much greater number of vertices (e.g. Unity, as discussed in \citealt{ferr} and also via private communication). Although UV mapping is a standard technique in 3D graphics, Blender offers two important advantages : 1) It is free and open source; 2) It has a Python interface - with Python increasing in popularity in the astronomy community, this can simplify the learning process for new users (see \citealt{bkent} and \citealt{nai}). 

Blender files containing Python scripts to load data in the manner described are available for download at the following URL : \href{http://tinyurl.com/jb8rf4n}{http://tinyurl.com/jb8rf4n}.

\section{Results}
\label{sec:res}
\noindent
Figure \ref{fig:conv} shows data displayed using a very simple transfer function, in which the data value sets both the opacity and colour of the mesh. Colour could also be uniform with only opacity allowed to vary. However, much more complex visualisation is possible. The data value can be used to simultaneously control both opacity and colour using different transfer functions for each (e.g. opacity can vary logarithmically while colour varies linearly). Opacity and colour do not have to be controlled by the same data set or even each use a unique data set - for example, from simulation data one could use density to set the opacity but temperature to control the colour. Multiple data sets can be overlaid using different transfer functions. In principle, one could even use multiple data sets to control the opacity and/or colour of the same mesh, allowing very detailed control over the display. Different colour schemes can also be used for different parts of the same data, enhancing contrast between specific regions - for an example of this see \cite{olivia} figure 1.

\begin{figure}[h]
\begin{center}
\includegraphics[width=85mm]{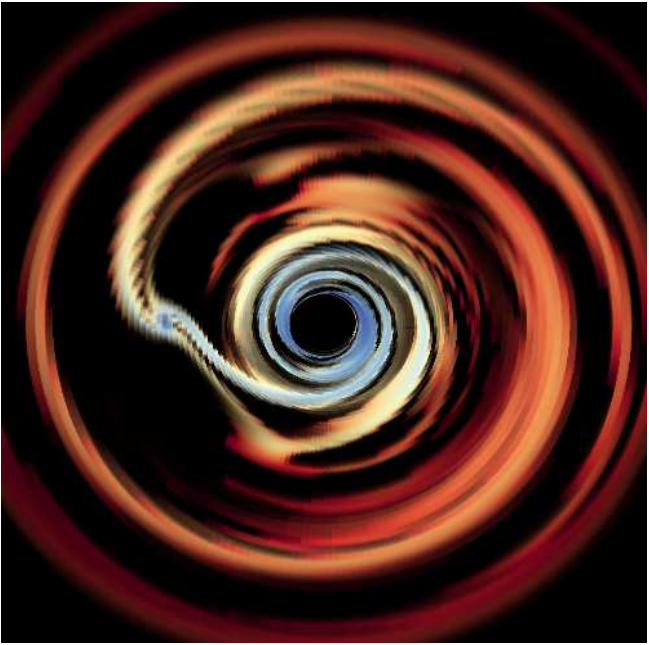}
\caption[DandQ]{Protoplanetary disc simulation using density to control the opacity and Q (the amount of heat energy generated in each cell) to set the colour (blue is high, red is low). A rotating movie is available at the following URL : \href{https://youtu.be/blbTWSp1cjI}{https://youtu.be/blbTWSp1cjI}.}
\label{fig:densq}
\end{center}
\end{figure}

Figure \ref{fig:densq} shows an example of a more complex visualisation techniques using the same data set as in figure \ref{fig:conv}. Here the opacity is controlled by the density while the colour is determined by the $Q$ parameter, a measure of how much heat energy is generated in each data cell. This is a particularly useful technique for simulation data, where one often has access to multiple data per pixel : instead of $Q$, one could use temperature, pressure, velocity or any other data. Velocity data can be hard to interpret using colour but it is also possible to plot velocity vector arrows.

\begin{figure}[h]
\begin{center}
\includegraphics[width=85mm]{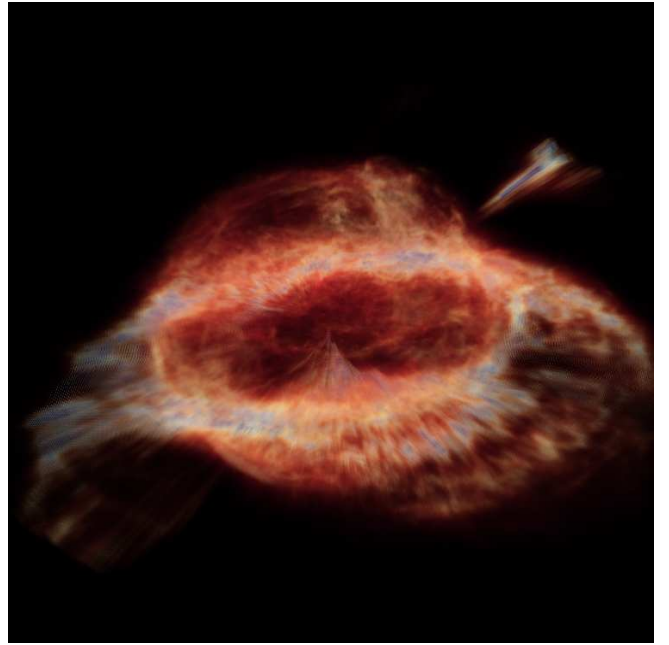}
\caption[LAB]{Visualisation of Leiden/Argentine/Bonn all -sky \HI{} survey data, using velocity for radial distance. Each velocity channel is mapped onto a separate sphere. Colour and opacity are controlled by intensity of the emission (red is low, blue is high). A movie is available at the following URL : \href{https://youtu.be/DMgXAKS1yI8}{https://youtu.be/DMgXAKS1yI8}.}
\label{fig:lab}
\end{center}
\end{figure}

One problem evident in the animations is that the data set disappears from certain viewing angles. In the standard version of \textsc{frelled}, the data are shown on planes - which disappear if viewed edge-on. The solution is to slice the data in three different projections, with the planes for each projection being orthogonal to the others. Thus at least one set of planes is always seen at high viewing angles. For spherical polar data the solution is a second set of spherical meshes, i.e. slices through the data at constant $r$ instead of $\theta$. An example of this is shown in figure \ref{fig:lab}, using data from the all-sky Leiden/Argentine/Bonn \HI{} survey (\citealt{lab}). Spheres require more vertices than cones, however as with the cones they do not require as many vertices as data points. In this example 320,000 vertices were used while the data contains 78 million voxels. The vertex count required is largely dependent on the number of spheres, i.e. the number of channels.

One other technique shown in the online version of figure \ref{fig:lab} is the use of varying the transfer function based on the data range in each slice. This was shown in T15 with Cartesian data but it is straightforward to adapt this to other data sets. Essentially, the same fraction of the data is displayed as completely transparent and opaque in each slice, rather than according to absolute data value. This method can greatly enhance the visibility of structures when the dynamic range varies depending on one particular coordinate, which can be useful for both observational and simulation data.

\section{Summary and Discussion}
\label{sec:disc}
\noindent
I have described a technique allowing the visualisation of 3D non-Cartesian data without needing to regrid that data. This means that the detail of the original data is perfectly preserved. Many of the visualisation techniques developed for \textsc{frelled} can be easily applied : multi-volume rendering, complex transfer functions and vector overlay are straightforward to adapt. Although the technique does require more GPU memory than for Cartesian data, this can be minimised to a level which is insignificant compared to the size of the data itself.

\begin{figure}[t!]
\begin{center}
\includegraphics[width=85mm]{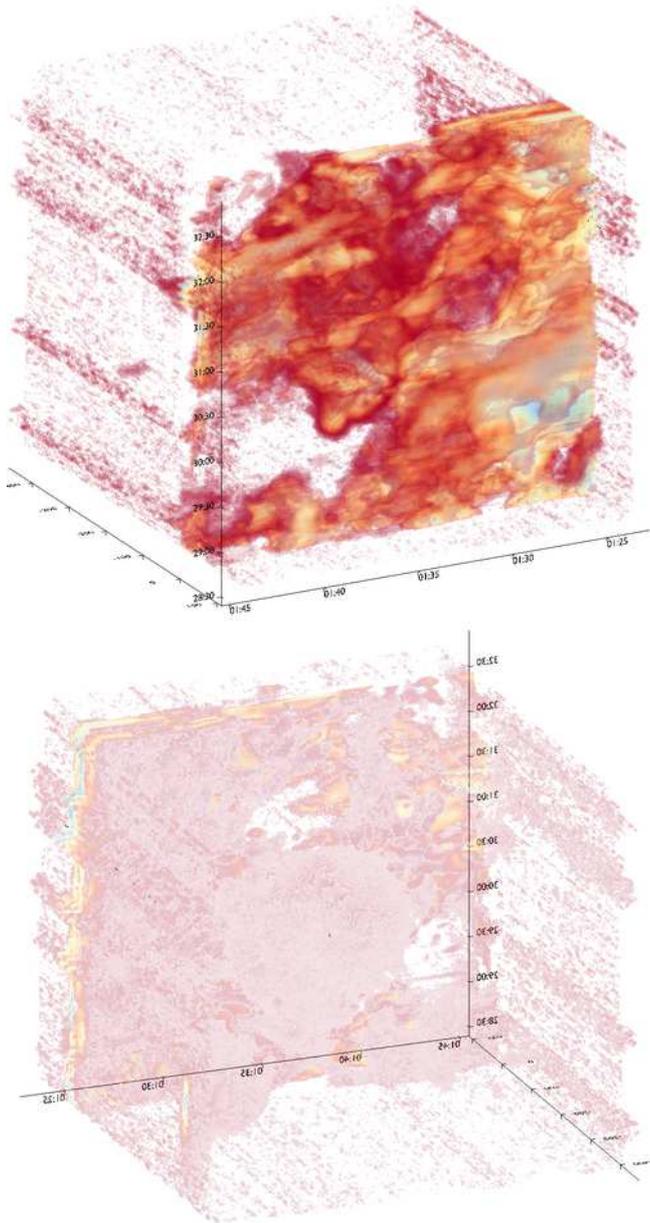}
\caption[M33]{Data cube visualised using \textsc{frelled} from two different sides. The top panel shows the cube from the side from which texture transparency is processed correctly in Blender. The lower panel shows the cube from the opposite side, showing obvious artifacts which are a limitation of Blender's realtime display engine. As described in the text there is a simple workaround for plane meshes, i.e. Cartesian data as shown here, but this is more complicated for data in spherical polar or arbitrary coordinates. The data set is the M33 cube described in \cite{olivia}. A rotating movie is available at the following URL : \href{https://youtu.be/CzcyMdbx9X4}{https://youtu.be/CzcyMdbx9X4}.}
\label{fig:M33ReverseAngle}
\end{center}
\end{figure}

However, considerable effort would be required to transform this into a fully-functioning FITS viewer, which is why the code presented here is only at the proof-of-concept stage. Firstly, \textsc{frelled} was developed with the primary aim of being a realtime viewer. One problem of this is that the realtime display engine in Blender only correctly displays transparent meshes from one side - from the other side the view is not useful, as shown in figure \ref{fig:M33ReverseAngle} and the accompanying animation. The solution in FRELLED is to use another set of meshes to display the other side (specifically, the process of \textit{duplicating} meshes in Blender forces a recalculation of the display such that the duplicated meshes are shown correctly from the opposite side as to the originals). A script running in the background constantly evaluates the angle of the view and chooses which set of meshes to display accordingly. 

This problem is relatively easy to address for plane meshes. Transparent spheres are more difficult, since the user is effectively seeing both sides of the mesh at once. Addressing this problem would require setting up a much more complicated series of meshes to ensure that everything is displayed correctly from every angle - possible, but non-trivial (creating meshes in Blender is too slow a procedure for the realtime view, so they must be pre-calculated). Currently the spherical code is really only useful for rendering static images and movies rather than for interactive content.

Selecting pixels within a specified region is also a potentially difficult problem. Many analysis tasks benefit greatly from being restricted to a particular region, e.g. a specific source. Objects in Blender can be of arbitrary shape, but automatically creating an object that matches the arbitrary coordinate system of the data - and especially for interactive scaling by the user - is not straightforward. External data sets are usually provided in Cartesian coordinates, so overlaying them would also have to correct for this difference. While challenging, none of these difficulties are insurmountable, making the method presented here a viable route for developing a viewer suitable for analysing data in arbitrary coordinate systems.

\section*{Acknowledgements}
\noindent
I am grateful to Hubert Klahr for providing the protoplanetary disc simulation data, and to Giles Ferrand for helpful discussions on visualising data with Unity. I also thank the anonymous reviewer whose comments improved the clarity of this manuscript.

This work was supported by the project RVO:67985815, the Czech Science Foundation project P209/12/1795 and the Tycho Brahe LG14013 project.

{}

\end{document}